# Statistical study of stacked/coupled site-controlled pyramidal quantum dots and their excitonic properties

S. T. Moroni, T. H. Chung, G. Juska, A. Gocalinska, E. Pelucchi

*Epitaxy and Physics of nanostructures, Tyndall National Institute, University College Cork, "Lee Maltings", Dyke Parade, Cork, Ireland*

**Abstract**

We report on stacked multiple quantum dots (QDs) formed inside inverted pyramidal recesses, which allow for the precise positioning of the QDs themselves. Specifically we fabricated double QDs with varying inter-dot distance and ensembles with more than two nominally highly symmetric QDs. For each, the effect of the interaction between QDs is studied by characterizing a large number of QDs through photoluminescence spectroscopy. A clear red-shift of the emission energy is observed together with a change in the orientation of its polarization, suggesting an increasing interaction between the QDs. Finally we show how stacked QDs can help influencing the charging of the excitonic complexes.

In the last decade semiconductor quantum dots (QDs) have been having a key role in the field of quantum optics and quantum information. Many challenges have been overcome over the years as semiconductor QDs have been proven to emit on-demand single, identical and entangled photons upon optical and electrical excitation [1][2][3][4][5]. Moreover, the atomic-like nature of the excitonic transitions in semiconductor QDs allowed performing fundamental studies on the complexes forming in a single-QD as such (e.g. [6][7][8]) and opened the way for a wider number of possible technological alternatives in the quest for quantum information processing.

One of the unique features of semiconductor QDs is the possibility to fabricate two QDs at a small distance in order to obtain a coupled QD system. Several studies were carried out on this kind of systems where the coupling of the electronic levels was obtained with different approaches, e.g. by lateral coupling [9] or the application of an electric field in order to match the electronic conduction level of the excitons and prepare a molecular-like state delocalized over two QDs [10] [11][12]. Nevertheless, most of the approaches to obtain coupled semiconductor QD systems are based on the vertical correlation of Stranski-Krastanov (SK) self-assembled QDs [10][13][14][15]. This methodology is, in general, intrinsically limited by a strong dot to dot "during growth" influence, delivering uneven dots and a broad statistical distribution of properties. Here, instead, we employ a different strategy to form stacked QDs, i.e. based on highly symmetric pyramidal site-controlled QDs [16]. Specifically, we took advantage of the uniformity and of the accurate control over the position of our pyramidal QD system to precisely stack two or more highly symmetric QDs on the top of each other. Remarkably, stacked coupling in technologically relevant site-controlled systems has been rarely addressed in the literature, with references [17] and [18] being the only cases known to us, both actually relying on "short" wire-like structures as QD systems as opposed to effectively self-assembled QDs.

Indeed, Pyramidal QDs are fabricated by performing MetalOrganic Vapour-Phase Epitaxy (MOVPE) over a GaAs substrate lithographically pre-patterned into an ordered array of pyramidal recesses. Inside each of the pyramids, an InGaAs QD layer is deposited between GaAs inner barriers and AlGaAs outer barriers. This QD family intrinsically delivers site-control and has been highlighted recently for its scalability potentiality when quantum technological approaches are to be considered. Actually, thanks to its intrinsic symmetry, this system has recently been proven to emit entangled photon pairs by means of both optical and electrical excitation [19][20]. In this work we filled the above-mentioned knowledge gap, we took advantage of a high level of control and reproducibility and addressed scalability issues by exploring the effects on QD formation and excitonic properties of various stacking recipes. This was done by collecting a large statistics on multiple-QD pyramids, characterizing InGaAs double-QD systems at different inter-dot barrier and higher number-QD pyramids, and strategically demonstrating a high level of control and tunability: an important characteristic for future exploitation as tailored quantum light sources [21].

We prepared a batch of four samples of pyramidal quantum dots composed of two stacked InGaAs QDs with the same nominal thickness (0.5 nm) and varying inter-dot GaAs barrier. The samples were grown on the same pre-patterned substrate to avoid any effect related to eventual small differences in the dimensions of the pyramidal recesses arising in separate substrates processing runs. All the samples were grown by MOVPE at 20mbar at a nominal temperature of 730 °C on a substrate patterned with 7.5μm pitch pyramidal recesses with growth conditions mimicking those reported elsewhere [22][23]. The different inter-dot barrier sizes were chosen to be 10 nm, 2 nm, 1 nm and 0.5 nm. The samples then underwent a conventional backetching procedure, consisting in the removal of the original substrate in order to turn the pyramids upside-down and reveal their tips [22][23]. All the samples were characterized

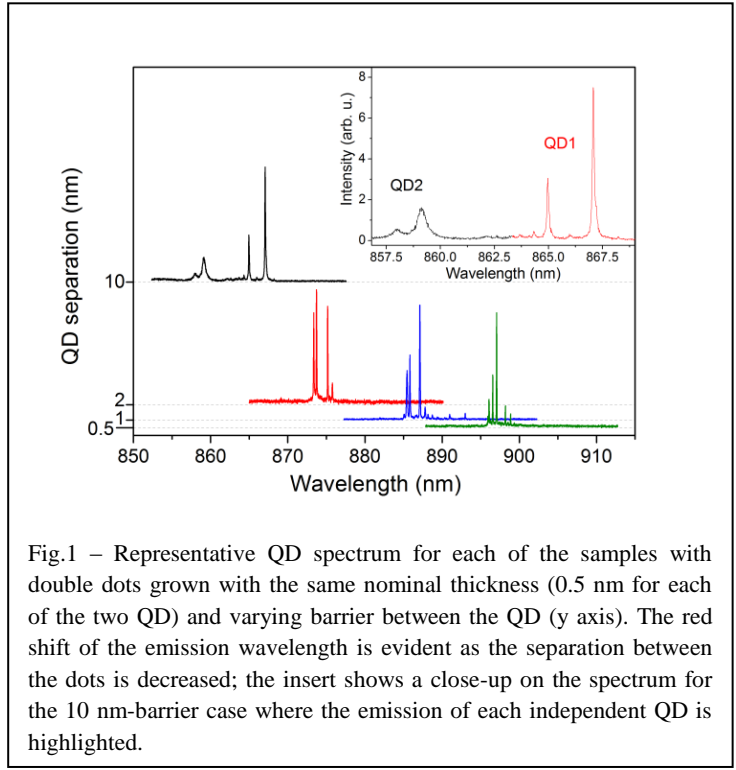

Fig.1 – Representative QD spectrum for each of the samples with double dots grown with the same nominal thickness (0.5 nm for each of the two QD) and varying barrier between the QD (y axis). The red shift of the emission wavelength is evident as the separation between the dots is decreased; the insert shows a close-up on the spectrum for the 10 nm-barrier case where the emission of each independent QD is highlighted.

by micro-photoluminescence spectroscopy at cryogenic temperature (~8 K).

Before starting our discussion, it should be highlighted that, although the QDs are nominally identical in thickness, the PL spectrum of each of them can be slightly different in emission energies and varies around a given mean value: this is due to unavoidable monolayer fluctuations in the thicknesses of the individual QDs as well as of the inter-dot barrier arising during the growth. Moreover, the second QD layer is deposited on the top of a surface profile that is slightly different from the GaAs profile on which the first QD is grown; therefore there may be minor differences in the shape of the two QDs as well [24][25].

A representative QD spectrum for each sample is reported in Fig.1. At first glance it is possible to see how the emission wavelength is red-shifting as the separation between the dots is reduced. A more detailed look at the spectra reveals that in the case of 10 nm spacing (Fig.1) typically it was always possible to distinguish two separate groups of emission peaks slightly (a few meV) shifted from each other, showing a similar pattern. Based on the comparison with previous data regarding single-dots [26], we were able to identify this peak pattern as the emission originated by the recombination of a negatively charged exciton and a biexciton for each of the two quantum dots. According to the picture we previously proposed in [26], the exciton recombination is completely hindered by a fast capture of one electron from the negatively charged surroundings, leading instead to what we identify as a negatively charged exciton transition. Interestingly, the 10 nm reference sample also systematically showed that the higher energy group of peaks has a broader linewidth than the more red-shifted one. We'll dedicate a more detailed description of this unexpected phenomenology in the following of our text and we concentrate, at this stage, our discussion on the other dot-separation samples.

Differently from the 10-nm-separation sample, the other three, which had an inter-dot barrier of 2 nm or lower, showed a single-QD-like emission pattern. By polarization-dependence, power dependence and cross-photon-correlation measurements (see supplementary material), for each sample we were able to identify only one transition associated to a single neutral exciton and one relative to a biexciton. A number of other (probably charged) transitions could be observed apart from neutral exciton and biexciton, but at the moment not enough information is available to fully describe their nature. Further studies (e.g. involving the application of an electric field) will help determining the origin and charging of these.

The change in behavior from two-QD-like to single-QD-like, together with the red-shift (a trivially expected behavior of dot coupling, but rarely reported with the here observed scale for SK dots for example; probably because SK processes deliver intrinsically non identical dots [27]) is suggesting that the stacked QDs are behaving like a single "molecular-like" QD rather than two independent single-QDs, even if it should not be considered per se as a full proof of an artificial molecule formation [18].

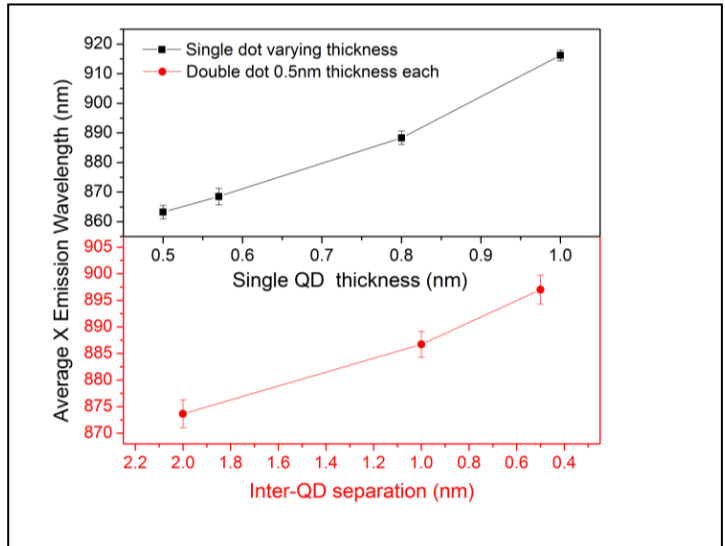

Fig.2 – Average emission wavelength as a function of the inter-QD barrier in stacked double 0.5 nm thickness QDs and comparison with the single-QD emission wavelength dependence on the QD thickness. About 20 double-QDs for each sample were considered for the averaging. The error bar shows the sigma of the emission dispersion for the measured dots.

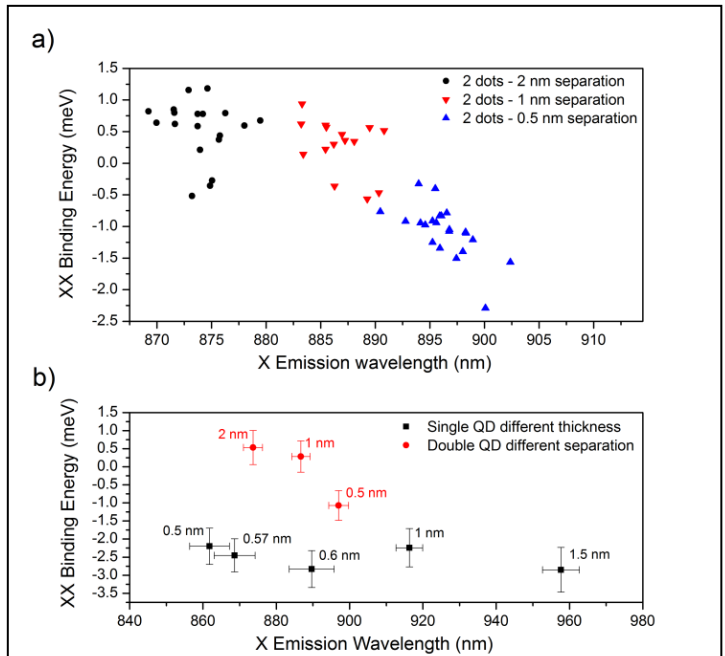

Fig.3 – a) Statistics for the biexciton binding energy as a function of the exciton emission wavelength across the different double QD samples; the inset shows the corresponding average emission wavelength for each sample. b) biexciton binding energy vs average exciton emission wavelength for different QD samples; the black squares are relative to single QD samples with different QD thicknesses (specified in the black labels) while the red dots indicate double 0.5 nm thickness QD samples with different separation (specified in the red labels).

To have a better insight on the effects of the QD stacking, a large statistics was collected for these three samples (about 20 QDs per sample type), taking into account measurements for the emission energy, the binding energy of the biexciton and the fine structure splitting for the exciton levels, known to be relatively small in PQDs. Fig.2 shows the average exciton emission wavelength for double QDs with 0.5 nm thickness and varying barrier thicknesses together with that for a typical single-QD with varying QD thicknesses. The average emission wavelength for the double QD samples is closer to that of a single QD with 0.5 nm thickness when the two QDs are distant (taking into consideration the usual 3-4 meV dispersion of the system [20][23]), and it varies to larger values to become closer to that of a single 1 nm-thick QD as the barrier is reduced. The biexciton binding energy statistics is plotted in Fig.3a. While an anti-binding biexciton with an average binding energy of about -2 meV is the typical case for a single pyramidal QD, for the stacked QDs we see that as the QD barrier is thickened the binding energy decreases and becomes positive (and the biexciton changes to binding). We take this as another sign of the fact that the two quantum dots are "communicating" and the charges composing the two electron-hole pairs are able to find an energetically favorable configuration (i.e. lower energy) when the QDs are kept at a larger distance rather than when they are very close to each other. It is clear from Fig.3a that a stronger correlation (statistically speaking) seems to have been established between the emission wavelength of the exciton and the binding energy of the biexciton as a function of inter-dot distance: the residual distribution can also be interpreted as an effect of the different influence (coupling) that one dot has on the other depending on (small) fluctuations in the thickness of the inter-dot barrier. Fig.3b is a comparison between the statistics obtained from the stacked-QDs with different barrier thickness (≤ 2 nm) and the regular single-QDs with different nominal thickness [26]: the switch in behaviour of the binding energy (never been observed as such in our single-QDs before) together with the difference in emission energy allows ruling out that the stacked-QDs with 2 nm, 1 nm and 0.5 nm separation could be in reality an elongated single-QD with 3 nm, 2 nm and 1.5 nm thickness respectively.

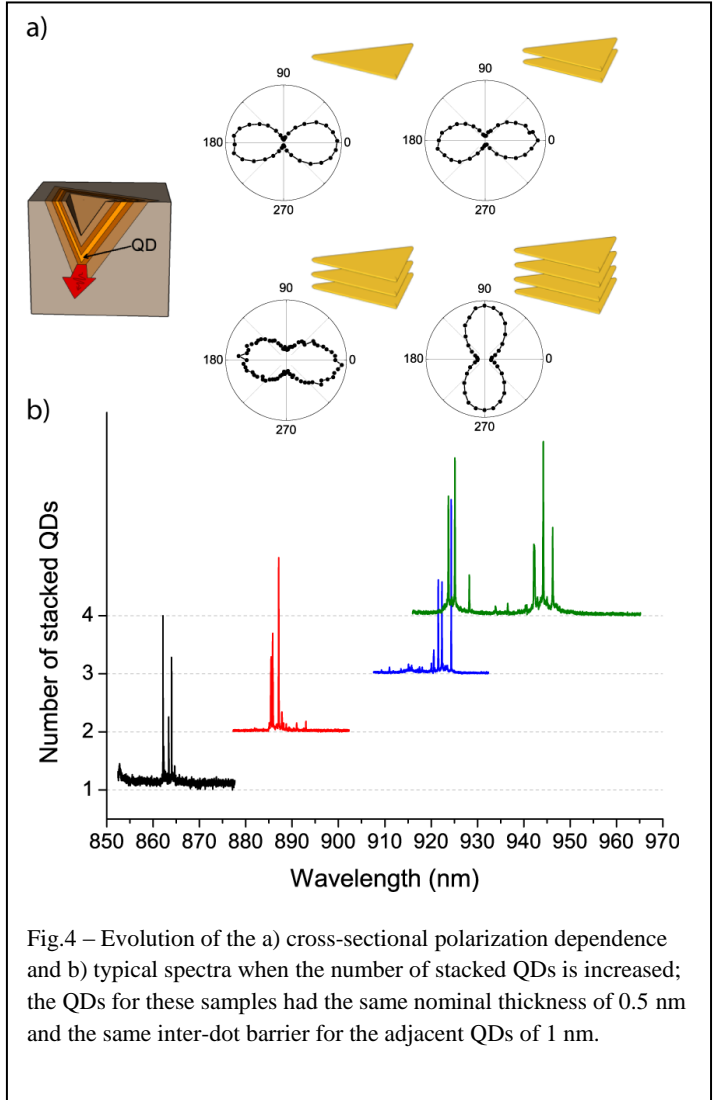

Fig.4 – Evolution of the a) cross-sectional polarization dependence and b) typical spectra when the number of stacked QDs is increased; the QDs for these samples had the same nominal thickness of 0.5 nm and the same inter-dot barrier for the adjacent QDs of 1 nm.

To improve our insight into the system, two more samples were grown with a different number of QDs: one with three 0.5nm-thick QDs and two 1nm-thick barriers in between and another sample with four 0.5 nm-thick QDs and 1nm-thick inter-QD barriers. A representative spectrum for each of the samples having one to four QDs and the same QD thickness (0.5nm) is presented in Fig.4. In each case a single-QD-like spectrum was measured. For each one it was possible to distinguish unambiguously the emission from the biexciton and exciton recombination cascade. The addition of a QD, again, causes the spectrum to red-shift and alters the binding energy of the biexciton bringing it to positive values. Cross-sectional polarization-dependence measurements ([6][28][29][30]) (i.e. measuring a cleaved facet of the sample, and not from the top) showed, surprisingly to some extent, that the polarization of the emission is oriented in the growth plane in the single-dot case and as the number of QDs is increased the luminescence starts showing a component oriented along the growth direction, becoming mainly aligned along the stacked QD axis in the case of four QDs (Fig.4), also a sign of electronic hybridization.

Interestingly, in the sample with four QDs, the increasing interaction between the dots also brings to the appearance of a set of higher energy transitions arising at higher excitation power. Systematic cross-correlations between these transitions and the lower energy ones lead to bunching in a number of cases, indicating that these transitions are related through a recombination cascade (see supplementary material). Although the pattern in the cascade and the relation between each peak is difficult to interpret and has not been fully understood yet, we might argue that these elements are perhaps hints of an even increasing "molecular" coupling between the QDs in the ensemble, including the possible appearance of "anti-bonding"- like states (which obviously are as today merely a possibility/speculation).

Before ending our letter, we want to finally discuss a different effect which is seen when QDs are stacked with a larger inter-dot barrier of 10 nm, as briefly mentioned earlier. The general behavior of the stacked double QD system in this case is that of two independent QDs, the one at higher energies showing always broader lines than the one at lower energies. To understand the source of this effect, we grew another sample with three QDs of *different thicknesses* and the same nominal inter-QD barrier of 10 nm, as reproduced schematically in Fig.5: the top and bottom QDs had a thickness of 0.45 nm while the central one was 0.6 nm thick. This allowed understanding which of the QDs in the ensemble (i.e. which one in growth order) has the best linewidth: as it is shown in the spectrum of Fig.5, the emission from the central dot can be easily distinguished as it is the lowest energy feature, while it can be a little more tricky to understand which of the other two dots the remaining emission belongs to.

It is well known that during the MOVPE growth of InGaAs on GaAs the lateral profile of the pyramidal recess tends to become larger [31]. Therefore we can speculate that the third grown QD (the bottom one, after the backetching process) will have a larger lateral profile, and, being grown at the same nominal thickness as the top one, it will actually result in being thicker. We conclude that most probably the top QD (i.e. the last grown one), which is closer to the surface after the backetching, is the one with broader lines (on a larger statistics of measured QD spectra, it shows an average value of 800± 400 μeV) while the bottom one shows the best linewidth (110± 30 μeV), the middle one having an intermediate broadening of the emission peaks (400±190 μeV). While the source of this systematic effect is not clear, we speculate that it could be related to the influence of the sample surface (which in all the samples presented here is only about 100nm away from the top dot) on the charge feeding of the QD. This is in agreement to the general trend observed in single (pyramidal) QDs where the presence of a negatively charged exciton leads to broader lines.

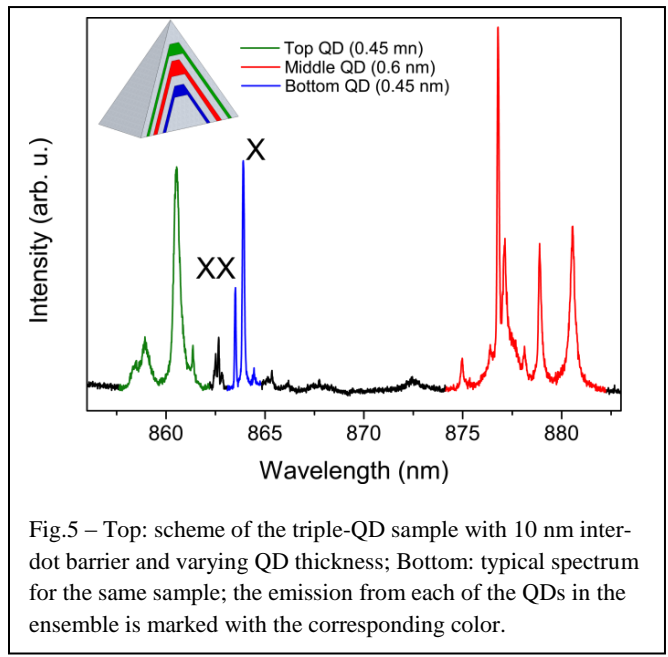

Fig.5 – Top: scheme of the triple-QD sample with 10 nm inter-dot barrier and varying QD thickness; Bottom: typical spectrum for the same sample; the emission from each of the QDs in the ensemble is marked with the corresponding color.

In conclusion, we investigated the effect that the stacking and interaction of two or more pyramidal highly symmetric quantum dots have on their optical properties. A consistent red-shift of the emission in identical stacked QDs and a change in the biexciton behavior were observed and interpreted as a sign of coupling between the QDs; this hypothesis was also validated by the evidence of a change in the polarization direction of the photoluminescence emission in higher-number stacked-QDs. Finally, we showed that the inclusion of an extra QD at a larger inter-dot separation from the first one can allow to systematically obtaining a better linewidth in above-bandgap excitation schemes. Our study also shows that the stacking of QDs can be an additional "tuning knob" to control the emission energy, biexciton binding energy, polarization and linewidth of our pyramidal QDs. Finally, more analysis will be carried out to prove and understand the nature of the coupling between the QDs and to confirm the appearance of anti-bonding states for multiple QWD coupling.

**Supplementary material**

See supplementary material for a comprehensive example of determination of exciton and biexciton transition in a QD molecule and for the details of the cross-correlation in a 4-stacked-QD sample.

**Acknowledgments**

This research was enabled by the Irish Higher Education Authority Program for Research in Third Level Institutions (2007-2011) via the INSPIRE programme, and by Science Foundation Ireland under grants 10/IN.1/I3000, 15/IA/2864 and 12/RC/2276. The authors are grateful to Dr K. Thomas for the MOVPE system support.